\documentclass{article}
\usepackage{spconf,amsmath,graphicx}

\usepackage{enumitem}
\setlist{nosep, leftmargin=14pt}

\usepackage{mwe} 

\title{Comparative Analysis of Unsupervised and Supervised Autoencoders for Nuclei Classification in Clear Cell Renal Cell Carcinoma Images}
\name{{\parbox[c]{\textwidth}{\centering Fatemeh Javadian$^{1}$, Zahra Aminparast$^{2}$, Johannes Stegmaier$^{1}$, Abin Jose$^{1}$}}}

\address{
 \small $^{1}$Institute of Imaging and Computer Vision, RWTH Aachen University, Aachen, Germany\\
 \small  $^{2}$Institute of Medical Sciences, Kermanshah University, Medical School, Kermanshah, Iran\\
}
\begin{document}
\ninept
\maketitle
\begin{abstract}
This study explores the application of supervised and unsupervised autoencoders (AEs) to automate nuclei classification in clear cell renal cell carcinoma (ccRCC) images, a critical diagnostic task that traditionally relies on subjective visual grading by pathologists. We evaluate various AE architectures, including standard AEs, contractive AEs (CAEs), and discriminative AEs (DAEs), as well as a classifier-based discriminative AE (CDAE), optimized using the hyperparameter tuning tool Optuna. Bhattacharyya distance is selected from several metrics to assess class separability within the latent space for different architectures, revealing challenges in distinguishing adjacent grades using fully unsupervised models. CDAE, integrating a supervised classifier branch, demonstrated superior performance in both latent space separation and classification accuracy.
Given that CDAE-CNN achieved notable improvements in classification metrics, affirming the importance of supervised learning for enhancing class-specific feature extraction in ccRCC nuclei, F1 score was incorporated into the hyperparameter tuning process to achieve the best-fitting model. The results demonstrate distinct improvements in identifying more aggressive ccRCC grades by leveraging the classification capability of AE through latent space clustering followed by supervised fine-grained classification. Our findings suggest that not only does the inclusion of a classifier branch in AEs provide a promising approach for improving grading automation in ccRCC pathology compared to unsupervised clustering or supervised learning alone, but the use of neural architecture search combined with contrastive learning techniques to structure embeddings in the latent space of the AE architecture also leads to enhanced performance, particularly in identifying more aggressive grades, which are more influential in tumor grading. This advancement has the potential to improve the accuracy of the final diagnosis.
\end{abstract}
\section{Introduction} \label{sec}

Grading nuclei in clear cell renal cell carcinoma (ccRCC), the most common subtype of renal cell carcinoma, is essential for accurate cancer diagnosis. Following World Health Organization (WHO) protocols, Whole Slide Images (WSIs) at 100× and 400× magnifications must be analyzed to assess nucleoli visibility for distinguishing grades 1 to 3 in ccRCC~\cite{1}. Higher grades generally indicate more aggressive cancer, requiring targeted treatments. Thus, deep learning models hold promise for facilitating diagnosis by reducing the labor-intensive and costly manual annotations involved in nuclei grading~\cite{2}. Artificial models can leverage detailed nuclear features to form similarity-based clusters, enabling grade prediction with minimal supervision~\cite{3}. Recent supervised methods employ high-resolution feature extraction networks to address the challenge of inter-class similarity in nuclei grading~\cite{4}. However, difficulties persist for fine-grained classification of ccRCC, especially in distinguishing closely related grades, such as grades 2 and 3, which are often confused due to subtle morphological differences~\cite{4}. While incorporating both image magnifications aligns with WHO protocol~\cite{5}, expert consultations indicate that experienced pathologists often rely primarily on 400× histopathological tissue analysis for accurate grading. This insight suggests the potential of using 400× magnification exclusively for nuclei classification.

Autoencoders (AEs)~\cite{6} have emerged as a promising technique for generating latent representations that capture essential nuclear features for classification~\cite{7}. AEs are applied for dimensionality reduction and latent space clustering, compressing high-dimensional nuclear data into embeddings that preserve critical features~\cite{8}. Additionally, the formation of clusters within this space corresponding to different nuclei grades has demonstrated effective feature retention~\cite{9}. Although contractive and discriminative modifications have been introduced to unsupervised AE models to enhance grade separation~\cite{10,11}, supervised and semi-supervised approaches further improve generalization by incorporating a classifier branch into the AE framework~\cite{12}.

This study aims to improve ccRCC classification across all grades, with an emphasis on the more aggressive grades that significantly influence diagnostic decisions. In clinical practice, the highest-grade nuclei often determine the overall grade of a WSI, directly impacting treatment choices~\cite{1}. We leverage AE's latent space clustering to separate images into four classes, evaluating the representation of target classes in an unsupervised latent space. Annotated data is then incorporated to refine separation based on grade-specific features, enhancing sensitivity to subtle morphological differences. Our results are compared with those from the Composite High-Resolution Network (CHR-Network)~\cite{4,5}, which shares the same dataset. Our hybrid method combines the clustering strengths of AEs with the benefits of supervised learning, aiming to outperform traditional models. Using a balanced dataset with limited annotations, we expect this approach to compensate for smaller training sets and reduce reliance on extensive manual annotations, making it practical for semi-supervised learning applications.

The methodology outlines model selection, network structure, and their impact on nuclei grading metrics. AE architectures are optimized with Optuna~\cite{13}, testing unsupervised and supervised models with distinct latent feature extraction methods. During data preparation, nuclei images are cropped to reduce background interference and emphasize differences. The experimental procedure includes data preparation, Optuna tuning, and results evaluation against a supervised model. Finally, we discuss the selected models and future work, including semi-supervised learning.

\section{Methodology}
\label{sec:format}

Two primary architectures, a Multi-Layer Perceptron (MLP) and a Convolutional Neural Network (CNN), were considered, each featuring a structured latent layer for efficient feature representation. The MLP uses a layered encoder and mirrored decoder, while the CNN employs sequential 2D convolutional layers with defined channels, kernel sizes, and activations, followed by a flattening layer and a dense layer leading to the latent space. The decoder reverses these layers, using transpose convolutions to reconstruct input. This modular structure enables rapid configuration testing with Optuna~\cite{13}, optimizing parameters across a diverse range of architectural and training hyperparameters. Beyond minimizing reconstruction loss, achieving clear class separation in the latent space is prioritized, with an additional metric evaluating representation quality to guide Optuna in optimizing classification.

Effective class separation in the latent space is assessed through cohesion and dispersion metrics. Cohesion, or intra-class similarity, measures how closely embeddings within a class cluster together, while dispersion, or inter-class separation, captures distances between class centroids~\cite{6}. The Davies-Bouldin Index~\cite{14}, Calinski-Harabasz Score~\cite{15}, and Silhouette Score~\cite{16} provide measures of class similarity, clustering quality, and class match, respectively. Additionally, MANOVA~\cite{17} analyzes means and variances across groups, though normality assumptions limit its use in latent spaces with non-Gaussian distributions.

While these metrics provide valuable insights into class structure, treating cohesion and dispersion as independent measures can lead to suboptimal interpretations, particularly in cases of high cohesion with limited inter-class separation~\cite{8,18}. To address this, Kullback-Leibler (KL) divergence~\cite{19} was considered for direct distributional divergence quantification, but due to its asymmetry and complexity in non-Gaussian spaces, Bhattacharyya distance~\cite{20} was ultimately selected as a symmetric and computationally efficient measure of distributional overlap.

The Bhattacharyya distance~\cite{20}, serves as a measure of separability between two probability distributions, quantifying their similarity by calculating overlap based on the Bhattacharyya coefficient. For two discrete distributions \(P\) and \(Q\), the Bhattacharyya coefficient (\(BC\)) is defined as:

\begin{equation}
      BC(P, Q) = \sum_{x} \sqrt{P(x) Q(x)}
\end{equation}
The Bhattacharyya distance \( D_B \) is derived as:
\begin{equation}
      D_B(P, Q) = -\log(BC(P, Q))
\end{equation}
This distance offers computational simplicity and symmetry, i.e., \( D_B(P, Q) = D_B(Q, P) \), making it ideal for latent space analysis without assuming specific distribution shapes.

To assess class separation in the latent space, dimensions are normalized using min-max scaling, and class histograms are created with bin counts based on the square root of the number of data points per class. These histograms are normalized to values within [0, 1], with entries scaled so their sum equals 1, allowing the histogram to serve as an approximate probability density. Pairwise Bhattacharyya distances are computed to form a symmetric distance matrix, bounded at 1. Overall class separability is evaluated by calculating the mean of the upper triangle of this matrix, excluding the diagonal. This mean quantifies the latent space’s capacity to distinguish between classes effectively, where larger values indicate a greater Bhattacharyya distance between different classes, signaling better separation.

After selecting the model structure and metrics, we next explored various AE types to optimize cluster separation and classification accuracy. Alongside the standard AE, two improved variants, contractive (CAE) and discriminative (DAE), as well as a supervised approach (CDAE), were evaluated.

CAEs, introduced by Rifai et al.~\cite{21}, reduce latent space sensitivity to minor input variations by incorporating the Jacobian matrix of the encoder’s output as an additional loss term. This Jacobian, measured with the Frobenius norm, quantifies input impact on embeddings in the latent space:

\begin{equation}
      \left\|J_f(x)\right\|_F^2=\sum_{i j}\left(\frac{\partial h_j(x)}{\partial x_i}\right)^2
      \label{eq:jacobian}
\end{equation}
where \(f\) is the encoder, \(J_f(x)\) is the Jacobian of \(f\), \(x\) the input, and \(i, j\) are indices iterating over input and latent space dimensions, respectively. To select the optimal contractive loss weight, Optuna is employed to scale the Frobenius norm of the Jacobian matrix~\cite{21}. This weight controls the impact of the contractive term in the CAE’s loss function, thereby reducing the sensitivity of the latent space to variations in input data.

DAEs~\cite{22} add a discriminative term to enhance class separation through cohesion and dispersion, using within-class (\(d_w\)) and between-class (\(d_b\)) losses:
\begin{equation}
      d_w = \frac{1}{n_{\mathbf{e}} \cdot n_{l}} \sum_{i=1}^{k} \sum_{\mathbf{e} \in C_{i}} ||\mathbf{e} - \mathbf{c}_{i}||^{2}
\end{equation}
\begin{equation}
      d_b = \frac{1}{k \cdot n_{l}} \sum_{i=1}^{k} \sum_{j=i+1}^{k} ||\mathbf{c}_{i} - \mathbf{c}_{j}||^{2}
\end{equation}
where \( \mathbf{e} \) is an embedding, \( \mathbf{c} \) the class centroid, \( k \) the class count, \( n_{\mathbf{e}} \) the number of embeddings, and \( n_{l} \) the latent dimension. The total discriminative loss \( l_d \) includes a balancing term:
\begin{equation}
      l_d = d_b + d_w + |d_b - d_w|
      \label{eq:discrimin}
\end{equation}
Here, centroids are model parameters, updated iteratively by:
\begin{equation}
      \mathbf{c}_i = \mathbf{c}_{i-1} - \alpha \cdot (\mathbf{c}_{i-1} - \mathbf{c}^{'}_{i})
\end{equation}
where \( \mathbf{c}^{'}_{i} \) is the current class \(i\) centroid in latent space, and \( \alpha \) is the update rate (set to 0.5). Optuna tunes the weight of the discriminative loss \( l_d \) (range 0.5 to 5) to balance class separation with other loss terms.

In the final step, a CDAE is created by adding a classification branch to the latent space. This branch consists of a dense layer that receives embeddings from the latent space, followed by a ReLU activation function, and an output layer with four nodes and LogSoftMax for generating classification outputs. The number of neurons in the dense layer is treated as a search dimension for Optuna, with values ranging from 4 to 32. To build the classification loss, negative log likelihood loss is used, which is then weighted and added to the reconstruction and discriminative losses. The classification branch weight is also optimized in the Optuna search space, with values from 1 to 10. By directly using class labels, the CDAE implements supervised learning, which is expected to outperform unsupervised models.

\begin{figure}[htb]
    \begin{minipage}[b]{1.0\linewidth}
      \centering
      \centerline{\includegraphics[width=8.5cm]{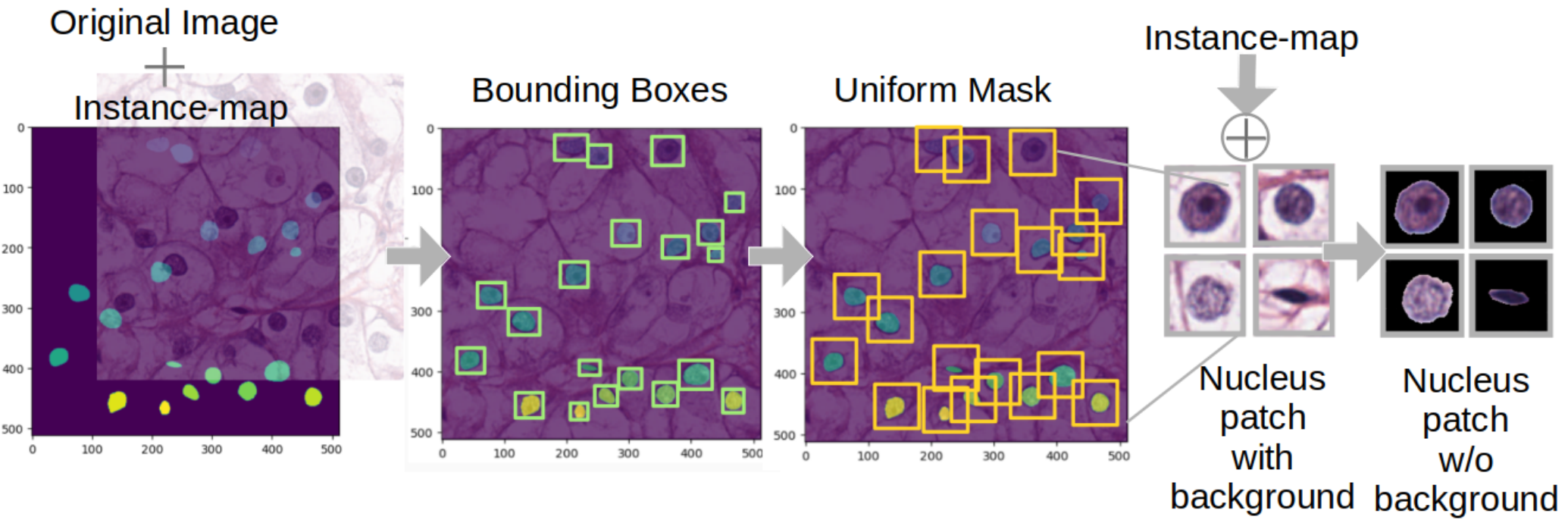}}
    \end{minipage}
    \vspace{-0.5cm} 
    \caption{\ninept Process of preparing nuclei patches for the AE. From left to right: nuclei are segmented using the instance map, enclosed within uniform bounding boxes, cropped, and then background is removed by reapplying the instance map to isolate nuclei within the patches.}
    \label{fig:image}
\end{figure}

\section{Experiments and Results}
\label{sec:pagestyle}

\subsection{Dataset Preparation}
The dataset comprises H\&E stained ccRCC images from TCGA, annotated by researchers as described in~\cite{4}. It includes 512×512 cropped images with nuclei annotations, covering segmentation and classification information for grades 1 through 3 and non-tumorous cells. These annotations are provided in \texttt{.mat} files, containing both instance and class maps. The instance map uniquely identifies each nucleus within clusters, while the class map assigns grade labels for classification. Data preparation involves segmenting each nucleus, labeling it by grade, and isolating it into patches based on the annotations. The background is removed using instance maps, isolating each nucleus in a separate patch while suppressing adjacent nuclei to focus on individual features, as illustrated in Figure~\ref{fig:image}. All nuclei patches are standardized to a uniform size for input into the autoencoder, with unannotated images used for grading in the classification extension. The original dataset includes 6,941 images in four classes: Grade 1 (3,782), Grade 2 (752), Grade 3 (850), and Non-tumorous (1,557); however, a balanced subset was used for training to mitigate dataset imbalance. Z-score normalization was applied using channel means and standard deviations derived from the training set, ensuring consistent scaling across validation and test sets.

\subsection{Experimental Procedure and Results}
\begin{figure}[htb]
\begin{minipage}[b]{.40\linewidth}
  \centering
  \centerline{\includegraphics[width=4.6cm]{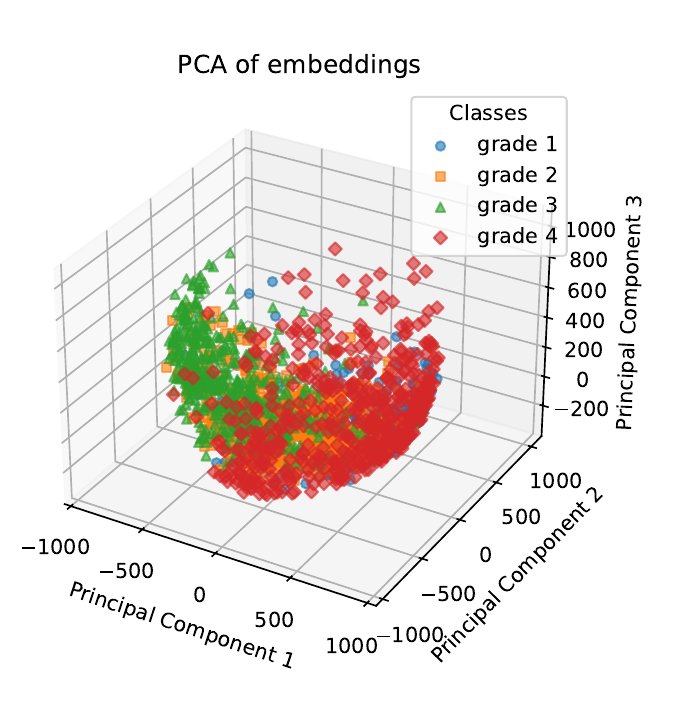}}
  \centerline{(a)}\medskip
\end{minipage}
\hfill
\begin{minipage}[b]{0.48\linewidth}
  \centering
  \centerline{\includegraphics[width=4.6cm]{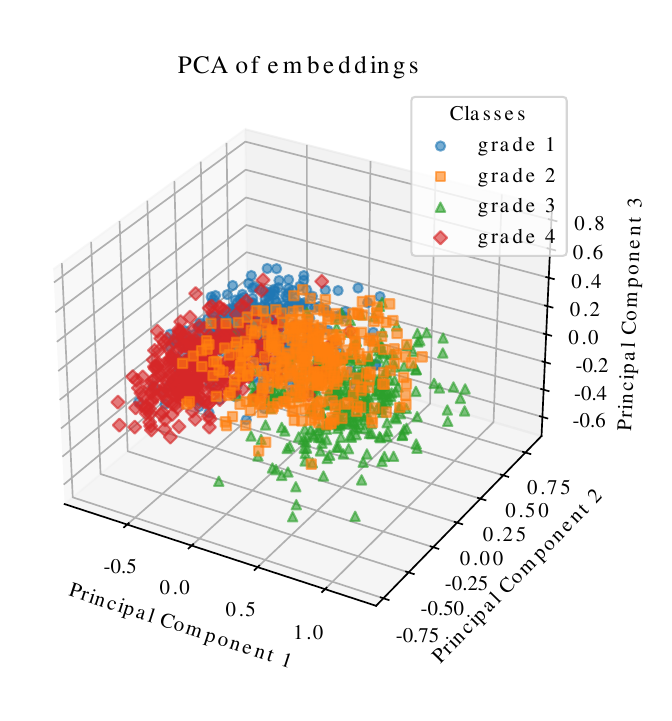}}
  \centerline{(b)}\medskip
\end{minipage}
\vspace{-0.5cm} 
\caption{\ninept Visualization of the first three PCA components for training results of the highest-performing latent space embeddings optimized for Bhattacharyya distance. (a) AE. (b) CDAE-CNN. Note that Grade 4 refers to Non-Tumorous in this context.}
\label{fig:good}
\end{figure}

Experiments were conducted on eight models—AE, CAE, DAE, and CDAE architectures, with MLP and CNN variants—on a balanced dataset with data augmentation through image flipping. The objective was to identify the optimal architecture by maximizing Bhattacharyya distance for cluster separation. Optuna was used to generate model configurations, with up to 1000 experimental trials employing a "funnel effect" to progressively reduce layer dimensions. For the standard AE, training used mean squared error (MSE) loss, with Bhattacharyya distance guiding optimization. MLP and CNN AEs achieved Bhattacharyya distances on the test split, as shown in Table~\ref{tab:distance}, but neither achieved clear class separation. A 3D visualization of PCA components from the CNN model optimized by Bhattacharyya distance is shown in Figure~\ref{fig:good} (a).

\begin{table}[h]
      \centering
      \ninept
      \caption{Bhattacharyya distance of highest-performing models found by Optuna after architecture search and fine tuning.}
      \label{tab:CDAE_f1}
      \begin{tabular}{rcccc}
            \hline
            Model & AE & CAE & DAE & CDAE \\
            \hline
            MLP   & 14.75    & 24.5    & 17.43    & 34.62    \\
            CNN   & 16.33    & 19.21    & 25.93    & 47.23    \\
            \hline
      \end{tabular}
      \label{tab:distance}
\end{table}
For CAEs, the contractive term enhanced latent space structure, as detailed in Table~\ref{tab:distance}. However, the contractive loss presented challenges, often causing latent space collapse and highlighting the need for careful weight selection to avoid over-compression. DAEs also improved latent space structure, with Bhattacharyya distances provided in Table~\ref{tab:distance}. Although not depicted here, PCA analyses from these models show slight improvement. However, despite these improvements, none of the unsupervised methods achieved satisfactory class separation. Consequently, CDAE models were investigated as a supervised approach. Following the same optimization procedure, CDAE-CNN architectures achieved strong separation, with Bhattacharyya distances detailed in Table~\ref{tab:distance}. For further analysis, a 3D visualization of the first three PCA components of training data embeddings from the CNN model optimized by Bhattacharyya distance is shown in Figure~\ref{fig:good} (b), leading to the selection of this structure for classification. Although Bhattacharyya distance was initially applied for neural architecture search and hyperparameter tuning, given that the F1 score is a key metric for assessing classification performance, we reapplied the neural architecture search to the CDAE-CNN model to optimize the overall F1 score. The selected CDAE-CNN model, optimized via F1, is identified as the best-performing model for the classification task. It uses convolutions with ReLU activation and batch normalization. The encoder consists of sequential convolutional layers with fine-tuned filter sizes (26, 23, 22, 17, 16, 15, 4, 3, and 2) and a uniform kernel size of 7, optimized during architecture search, with each layer having a fixed stride of 1. These layers are followed by a flattening layer, a dense layer of 7 neurons with ReLU activation, leading into a latent space of dimension 10. The decoder mirrors this structure to reconstruct the input. For classification, the latent representation is passed through an 18-neuron dense layer with ReLU activation, followed by an output layer with 4 neurons and LogSoftMax activation. Classification loss uses a weighted negative log likelihood, combined with reconstruction and discriminative losses, with the weight of the classification branch optimized by Optuna. The model takes an identical input size, producing a matching output size in the decoder, while the classification branch outputs a vector of length 4. To ensure reliability, the best-performing model was retrained multiple times with the same hyperparameters to investigate the deviations and average performance metrics. The optimal configurations for both metrics are analyzed for improvements in overall and per-class performance, as presented in Table~\ref{tab:pm}. When optimized by F1, the CDAE-CNN model showed improvements in precision, recall, and F1 across all classes, despite a lower Bhattacharyya distance. Figure~\ref{fig:good} illustrates a comparison of latent space distributions for both models, indicating superior class separation in the model optimized by Bhattacharyya distance compared to the F1-optimized model with lower Bhattacharyya distance value but better classification performance metrics. Although we demonstrated that supervised AEs outperform unsupervised methods in ccRCC nuclei grading, it is still necessary to assess if this model surpasses other supervised approaches, especially in separating adjacent and more critical grades 2 and 3. Therefore, we compared our results with those from CHR-Network~\cite{4,5}, specifically due to its fine-grained classification approach and sharing the original dataset with our study. The evaluation metrics, shown in Table~\ref{tab:performance_metrics}, indicate that our model outperforms the CHR-Network in all parameters except for grade 1. Notably, we used a balanced dataset, while the CHR-Network used an unbalanced dataset with counts of 45,108, 6,406, 2,779, and 16,652 for grades 1 to 3 and non-tumorous cells, respectively. The dominance of grade 1 in this unbalanced dataset skews overall performance, complicating direct comparisons. To address this, we applied balanced accuracy~\cite{23} to mitigate the effect of dataset imbalance on CHR-Network performance. The results in Table~\ref{tab:performance_metrics} confirm that the AE approach indeed enhances cluster separation and grading accuracy, particularly for more challenging grades. This underscores that leveraging AEs alongside supervised learning can outperform models relying solely on supervised learning.

\begin{table}
    \centering
    \ninept
    \caption{Comparison of Optuna optimization results using Bhattacharyya distance versus F1 score.}
    \label{tab:pm}
    \begin{tabular}{p{3cm}ccc}
        \toprule
        Metric & Bhattacharyya Opt. & F1 Opt. \\
        \midrule
        Overall Precision & 0.6104 & \textbf{0.6985} \\
        Overall Recall & 0.6111 & \textbf{0.7008} \\
        Overall F1 & 0.6113 & \textbf{0.6994} \\
        Grade 1 F1 & 0.5571 & \textbf{0.6373} \\
        Grade 2 F1 & 0.4469 & \textbf{0.5481} \\
        Grade 3 F1 & 0.6890 & \textbf{0.7821} \\
        Non-tumorous F1 & 0.7520 & \textbf{0.8300} \\
        Bhattacharyya distance & \textbf{47.23} & 36.90 \\
    \bottomrule
    \end{tabular}
\end{table}

\begin{figure}[htb]
\begin{minipage}[b]{.40\linewidth}
  \centering
  \centerline{\includegraphics[width=4.5cm]{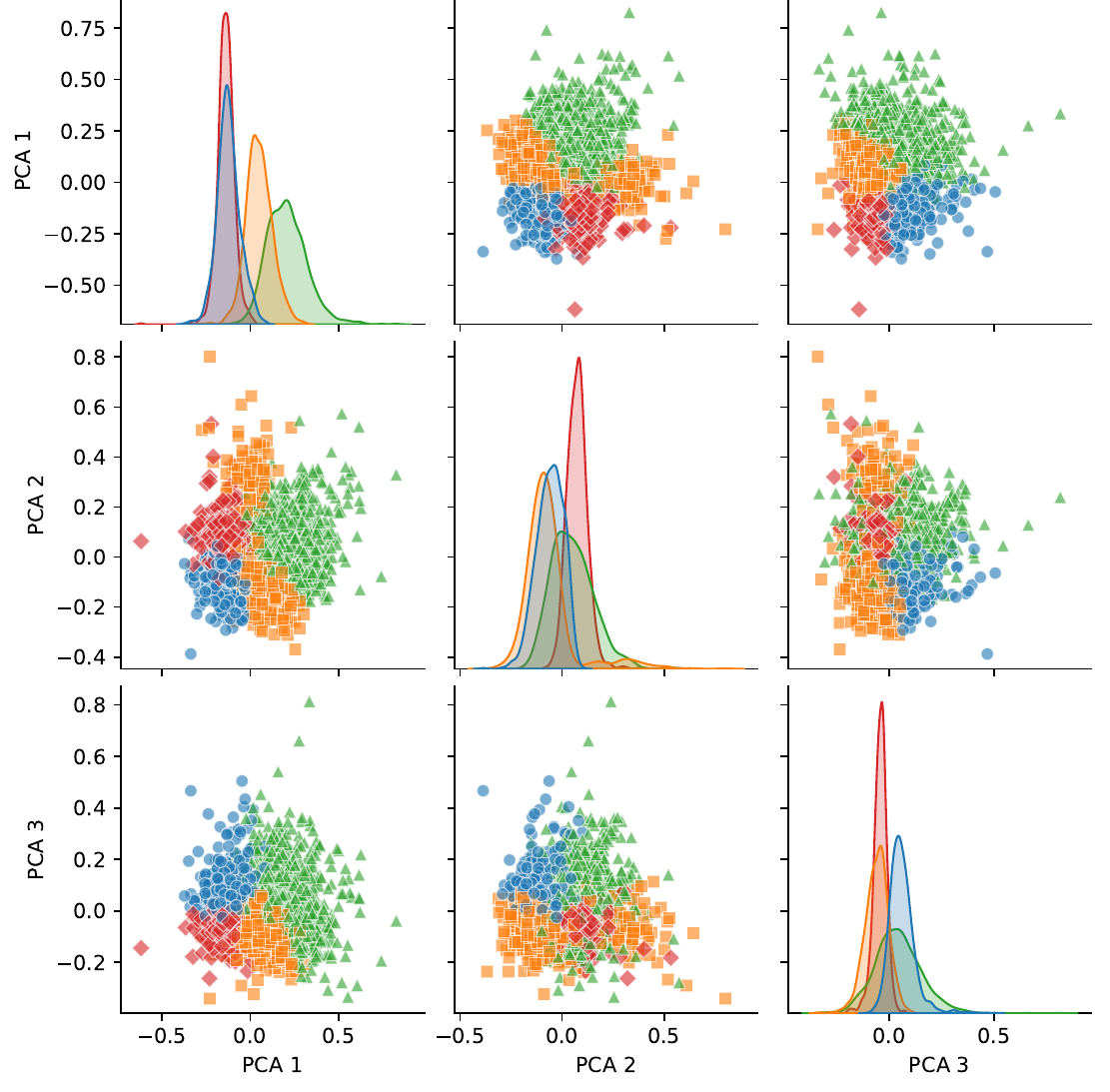}}
  \centerline{(a)}\medskip
\end{minipage}
\hfill
\begin{minipage}[b]{0.48\linewidth}
  \centering
  \centerline{\includegraphics[width=4.8cm]{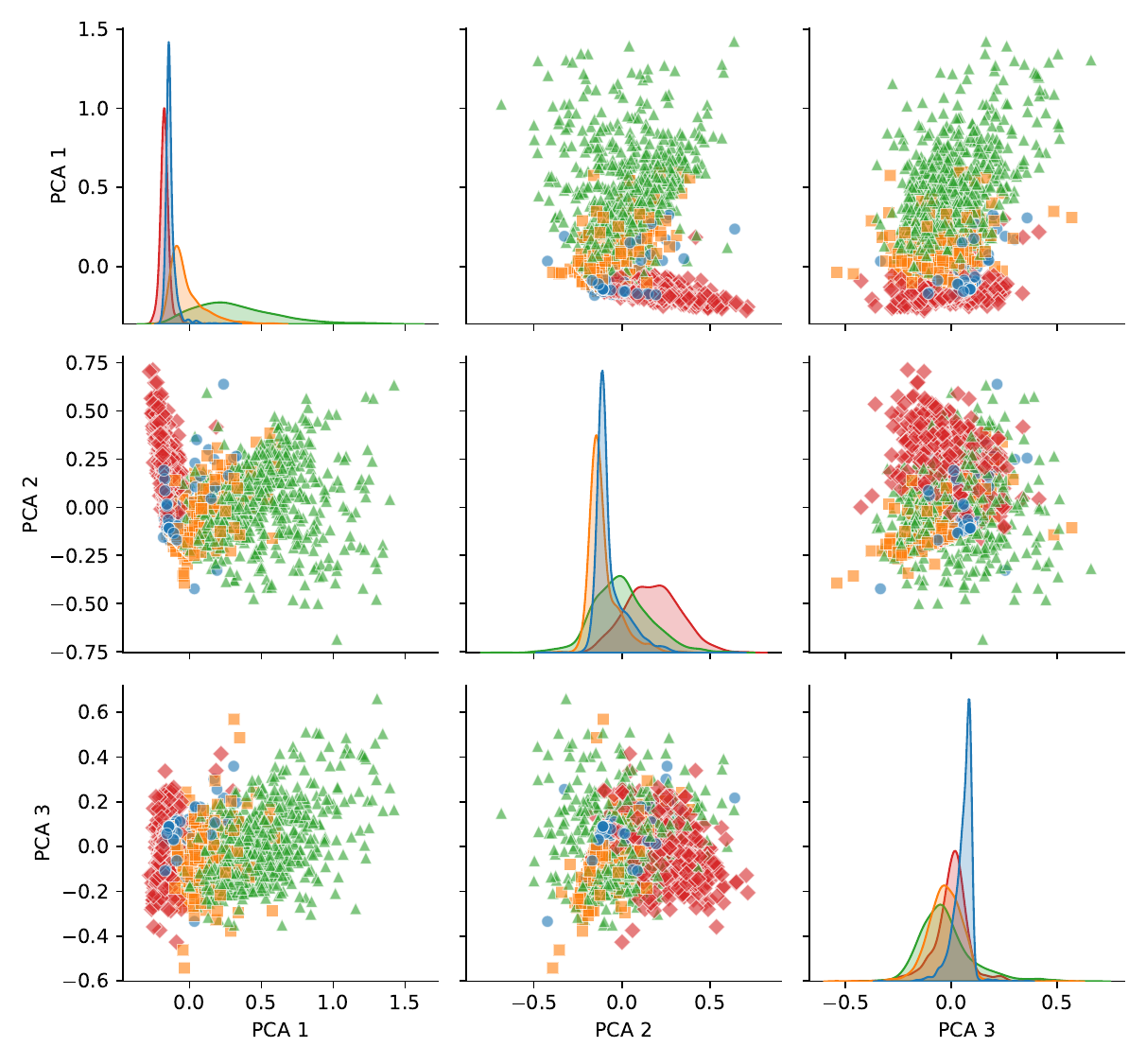}}
  \centerline{(b)}\medskip
\end{minipage}
\vspace{-0.5cm} 
\caption{\ninept Pair plots of the first PCA elements for the highest performance AE latent space embeddings. (a) The train results of CDAE-CNN for the highest performance model optimized for Bhattacharyya distance. (b) The train results of CDAE-CNN for the highest performance model optimized for F1-score.}
\label{fig:res}
\end{figure}
\begin{table}
    \centering
    \ninept
    \caption{Comparison of classification metrics for supervised learning results of CHR-Net and the selected CDAE model.}
    \label{tab:performance_metrics}
    \begin{tabular}{rcccc}
        \toprule
        Metric & CHR-Net  & CDAE \\
        \midrule
        Grade 1 F1 & \textbf{0.8243} & 0.6373 \\
        Grade 2 F1 & 0.4793 & \textbf{0.5481}\\
        Grade 3 F1 & 0.5228 & \textbf{0.7821} \\
        Non-tumorous F1 & 0.7057 & \textbf{0.8300} \\
        Balanced Accuracy & 0.640475 & \textbf{0.7008} \\
        \bottomrule
    \end{tabular}
\end{table}
\section{Discussion, Conclusion and Outlook}
\label{sec:typestyle}

The results show that standard AE models yield the lowest performance across CNN and MLP architectures (Table~\ref{tab:distance}). While structuring the latent space in CAE improves clustering in the CNN model and the DAE shows enhancement with MLP (Table~\ref{tab:distance}), these still fall short of performance goals. Fully unsupervised clustering of ccRCC nuclei, based solely on intrinsic features, does not consistently achieve the desired class separation. Overlaps at grade boundaries add complexity, often requiring subjective interpretation by pathologists~\cite{24}, resulting in a stochastic classification process with inherent uncertainty. By adding a classifier branch, the CDAE model surpasses others in latent space separation (Table~\ref{tab:distance}) and uses feedback from annotated data for clearer class differentiation. Consequently, the CDAE-CNN architecture, achieving the highest Bhattacharyya distance, was selected with the best discriminating latent structuring (Table~\ref{tab:distance}). Figure~\ref{fig:good} compares unsupervised and supervised AEs, highlighting the supervised model’s superior clustering and class boundary definition. This enhanced clustering pattern in the CDAE shows how labeled data better enables the model to capture subtle morphological differences between nuclei grades. While a higher Bhattacharyya distance indicates improved cluster separation, the F1 score directly reflects classification performance. A second Optuna search using F1 score rather than Bhattacharyya distance assessed how this alternative metric impacts outcomes. As shown in Table~\ref{tab:pm}, the CDAE-CNN model optimized by F1 achieves enhanced classification despite slightly reduced cluster separation. This comparison underscores the impact of metric selection on neural architecture search and final performance. A key result was the comparison with the CHR-Network. Although our study removed the background and analyzed nuclei in isolation, omitting information about distribution and positioning, our supervised AE approach still outperformed the CHR-Network in separating and classifying nuclei grades. Furthermore, our training dataset was smaller, reducing dependency on large labeled datasets. While neural architecture search may be computationally intensive, this limitation is mitigated in clinical tasks, as it is unnecessary to apply the search for every modification. Instead, an optimized model can be chosen for each application, with hyperparameter fine-tuning applied only when needed.

In conclusion, this study shows that integrating a supervised classifier branch significantly improves model performance for classification tasks where data features do not naturally align with target clusters. This approach combines the dimensionality reduction strength of autoencoders with the classification capabilities of supervised learning, enhancing fine-grained ccRCC nuclei grading and diagnostic precision. In future work, we aim to extend this method to a semi-supervised learning approach, reducing reliance on manual annotations by leveraging a larger pool of unlabeled data for pretraining and fine-tuning on a smaller labeled dataset.

\section{Compliance with Ethical Standards}
\label{sec:com}
This study utilized publicly available data, and therefore, no approval from an ethics committee was required.

\bibliographystyle{IEEEbib}
\bibliography{strings,refs}

\end{document}